\documentstyle[12pt]{article}
\def\beq{\begin{equation}}
\def\eeq{\end{equation}}
\def\bea{\begin{eqnarray}}
\def\eea{\end{eqnarray}}
\def\bq{\begin{quote}}
\def\eq{\end{quote}}

\def\gappeq{\mathrel{\rlap {\raise.5ex\hbox{$>$}}
{\lower.5ex\hbox{$\sim$}}}}

\def\lappeq{\mathrel{\rlap{\raise.5ex\hbox{$<$}}
{\lower.5ex\hbox{$\sim$}}}}

\begin{document}
\topmargin -0.5cm
\oddsidemargin -0.3cm
\pagestyle{empty}
\begin{flushright}
{HIP-1998-61/TH}
\end{flushright}
\vspace*{5mm}
\begin{center}
{\bf The generalisation of the Coulomb gauge to Yang-Mills theory} \\
\vspace*{1.5cm} 
{\bf Christofer Cronstr\"{o}m}$^{*)}$ \\
\vspace{0.3cm}
Physics Department, Theoretical Physics Division  \\
FIN-00014 University of Helsinki, Finland \\
\vspace{0.5cm}
 
\vspace*{3cm}  
{\bf ABSTRACT} \\ \end{center}
\vspace*{5mm}
\noindent
I consider the problem of generalising the Abelian Coulomb gauge condition to the non-Abelian 
Yang-Mills theory, with an arbitrary compact and semi-simple gauge group. It is shown that a 
straightforward generalisation exists, which reduces the Gauss law into a form involving the 
gauge potentials only, but not their time derivatives. The existence and uniqueness of the 
generalised  Coulomb gauge is shown to depend on an elliptic linear partial differential equation 
for a Lie-algebra valued quantity, which defines the gauge transform by means of which the 
generalised Coulomb gauge condition is realised. Thus the Gribov problem is actually non-existent 
in this case.

\vspace*{1cm} 
\noindent 
11.15.-q, 04.20.Fy, 04.60.Ds 

\vspace*{2cm} 
\noindent 
$^{*)}$ e-mail address: Christofer.Cronstrom@Helsinki.fi.
\vspace*{1cm}
\begin{flushleft}
September 12, 1998
\end{flushleft}
\vfill\eject


\setcounter{page}{1}
\pagestyle{plain}

\section{Notation and Conventions}  

The basic variables in Yang-Mills theory \cite{YM} are the gauge field $G_{\mu \nu}$
and gauge potential $A_{\mu}$, respectively. These quantities take values in  a convenient 
(Hermitian) matrix representation  of  the Lie algebra of the gauge group G. The Lie algebra is 
defined by the structure constants $f_{ab}^{~~c}$ in the  commutator algebra of the Hermitian 
matrix representatives $T_{a}$ of the Lie algebra generators,
\begin{equation}
[T_{a}, T_{b}] = if_{ab}^{~~c}T_{c},
\label{eq:liealg}
\end{equation}
where appropriate summation over repeated Lie algebra indices $(a, b, c, ..., f)$ is understood.
We assume $G$  to be semisimple and compact. Then a  positive definite Lie algebra metric 
can be given in terms of the following Killing form $(h_{ab})$,
\begin{equation}
h_{ab} = -f_{ab'}^{~~c'}f_{bc'}^{~~b'}
\label{eq	:Killing}
\end{equation}
The inner product $(A, B)$ of any two Lie algebra valued quantites $A = A^{a}T_{a}$ and
$B = B^{a}T_{a}$ is defined as follows,
\begin{equation}
(A, B) = h_{ab}A^{a}B^{b}
\label{eq:inprod}
\end{equation}
The form $h_{ab}$ and its inverse $h^{ab}$ are used to lower and raise Lie algebra indices, 
respectively. 

In the notation introduced so far, we write the gauge potential $A_{\mu}$ as follows,
\begin{equation}
A_{\mu}(x) = A_{\mu}^{a}(x) T_{a}
\label{eq:potA}
\end{equation}
where the argument $x$ stands for a space-time point in Minkowski space, which will be used as
a base space. The gauge field $G_{\mu\nu}(x)$ is then given as follows,
\begin{equation}
G_{\mu\nu}(x) = \partial_{\nu}A_{\mu}(x) - \partial_{\mu}A_{\nu}(x) - ig[A_{\mu}(x), A_{\nu}(x)],
\label{eq:Gfield}
\end{equation}
where $g$ is an arbitrary nonvanishing real parameter, which is introduced for convenience.
An alternative to the notation (\ref{eq:Gfield}) is the following,
\begin{equation}
G_{\mu\nu}^{a}(x) = \partial_{\nu}A_{\mu}^{a}(x) - \partial_{\mu}A_{\nu}^{a}(x) + gf_{bc}^{~~a}
A_{\mu}^{b}(x), A_{\nu}^{c}(x)
\label{eq:Gindfield}
\end{equation}

Our remaining notation is fairly conventional. Greek letters $\mu, \nu, ...$ are spacetime indices 
which take values in the range $(0,1,2,3)$. These indices are lowered (raised) with the standard 
diagonal Minkowski metric $g_{\mu\nu}\; (g^{\mu\nu})$ with signature $(+,-,-,-)$. Latin indices 
from the middle of the alphabet $(k, \ell, ...)$ are used as space indices in the range $(1, 2, 3)$. 
Unless otherwise stated, repeated indices are always summed over, be they Lie algebra-, spacetime- 
or space indices.

We finally state the gauge transformation formulae for the gauge potential and field, respectively,
in our notation. Let $\Omega(x)$ denote a general gauge transformation, which here may be 
identified with a unitary matrix of the following form,
\begin{equation}
\Omega(x) = \exp(ig\alpha^{a}(x) T_{a}),
\label{eq:gaugetrans}
\end{equation}
where the functions $\alpha^{a}(x)$ are any sufficiently smooth real-valued  functions on Minkowski 
space. The relevant gauge transformation formulae are as follows,
\begin{equation}
A_{\mu}(x)  \stackrel{\Omega}{\longrightarrow} A'_{\mu}(x) = \Omega^{-1}(x) A_{\mu}(x) \Omega(x)
+ \frac{i}{g} (\partial_{\mu}\Omega^{-1}(x))\Omega(x),
\label{eq:gtrA}
\end{equation}
and
\begin{equation}
G_{\mu\nu}(x) \stackrel{\Omega}{\longrightarrow} G'_{\mu\nu}(x) = \Omega^{-1}(x) G_{\mu\nu}(x) \Omega(x)
\label{eq:gtrG}
\end{equation}

After these preliminaries the new generalised Coulomb gauge condition will be given and discussed 
in the next section.

\section{The generalised Coulomb gauge condition}
 Let us first recall the usual Coulomb gauge condition in electrodynamics,
\begin{equation}
\nabla \cdot {\bf A}(x^{0}, {\bf x}) \equiv \partial_{k}A^{k}(x^{0}, {\bf x}) = 0
\label{eq:Coul}
\end{equation}
The motivation for using condition (\ref{eq:Coul}) in electrodynamics, both 
classical and quantum, is sound; it enables one to solve Gauss law explicitly
for the time component of the potential $A_{0}$ and leads to a fairly satisfactory 
canonical formalism, as is well known. The lack of manifest covariance of condition 
(\ref{eq:Coul}) under Poincar\'{e} transformations is not in any way serious; this 
circumstance can be compensated for by adjoining an appropriate gauge transformation to
the Lorentz-transformation of the potential $A_{\mu}$.

In pioneering papers \cite{Schw 1}, \cite{Schw 2} dealing with the quantization of Yang-Mills theory   
the condition (\ref{eq:Coul}) was taken over {\em verbatim} and was subsequently generally 
accepted  as a {\em bona fide} gauge condition also in non-Abelian gauge theory until V. N. 
Gribov in 1977 demonstrated that condition (\ref{eq:Coul}) strictly speaking is not a gauge 
condition at all in the non-Abelian case \cite{Gribov}. It was later shown by Singer 
\cite{Singer} that the so-called Gribov ambiguity persists for {\em any} gauge condition,
if the gauge potentials are defined on compact versions of space-time such as e.g. $S_{3}$ or
$S_{4}$, or stated more generally, if the base space is a compact manifold. For potentials
defined on non-compact Minkowski space, the compactness condition can be taken to mean, 
roughly speaking, that the gauge potential and  gauge field, respectively, ought to vanish 
sufficiently rapidly at infinity. The fact that there exists at least one gauge condition in 
non-Abelian gauge theory, which does not suffer from the Gribov ambiguity, namely the so-called 
Fock-Schwinger gauge \cite{FS}, then indicates that one must be prepared to accept gauge potentials 
and gauge fields with non-trivial asymptotic behaviour, so that the compactness assumptions
in Singer's general argument do not apply.

After all this I will now write down the generalised Coulomb gauge condition advertised in the title
of this paper. The condition in question is simply the following,
\begin{equation}
\nabla_{k}(A) {\partial_{0}A^{k}}(x) \equiv \partial_{k} \partial_{0}A^{k}(x) +
ig[A_{k}(x), \partial_{0}A^{k}(x)] = 0
\label{eq:genCoul}
\end{equation}

It is clear that condition (\ref{eq:genCoul}) reduces to the ordinary Coulomb gauge condition
(\ref{eq:genCoul}) differentiated with respect to time $x^{0}$ in the Abelian case, when the 
commutator term in (\ref{eq:genCoul}) disappears. However, it is {\em a priori} not clear
that condition (\ref{eq:genCoul}) in the general non-Abelian case {\em is} a gauge condition; 
that this is in fact the case will be demonstrated presently.  The motivation for condition
(\ref{eq:genCoul}) comes from my analysis  of the canonical structure of Yang-Mills theory. 
This will be briefly touched upon subsequently, and presented in greater detail in 
a separate paper \cite{Chris 1}.

I will now demonstrate that condition (\ref{eq:genCoul}) is a gauge condition. It will be convenient 
to denote the time derivative of any quantity with a dot on top of that quantity, thus for example
\begin{equation}
\dot{A}_{k}(x) \equiv \partial_{0}A_{k}(x)
\label{eq:dot}
\end{equation}
Likewise it is convenient to introduce the notion "covariant gradient" $\nabla(A)$ for the 
differential operator operating on $\dot{A}^{k}(x)$ in (\ref{eq:genCoul}), i.e.
\begin{equation}
\nabla_{k}(A) \equiv \partial_{k} + ig[A_{k}(x), \;\;]
\label{eq:covgrad}
\end{equation}
To demonstrate that (\ref{eq:genCoul}) is a gauge condition, is equivalent to showing that there 
exists a gauge transformation $\omega$, say, by means of which a general gauge potential 
${\cal A}_{\mu}$, say, not necessarily satisfying any particular gauge condition, gets 
gauge transformed into a potential $A_{\mu}$ satisfying condition (\ref{eq:genCoul}). Hence 
we consider the relation
\begin{equation}
A_{k}(x) = \omega^{-1}(x) {\cal A}_{k}(x) \omega(x) + \frac{i}{g}(\partial_{k}\omega^{-1}(x))\omega(x)
\label{eq:specA}
\end{equation}
It is now convenient to define a quantity $X_{0}$ as follows,
\begin{equation}
X_{0}(x) \equiv  -\frac{i}{g}\,{\dot {\omega}(x)} \omega^{-1}(x)
\label{eq:capX}
\end{equation}
It is clear that if the quantiy $\omega(x)$ above takes values in  the gauge group, then the 
quantity $X_{0}$ defined in Eq. (\ref{eq:capX}) takes values in the corresponding Lie algebra.

Differentiating Eq.(\ref{eq:specA}) with respect time $x^{0}$ one readily obtains the following
result,
\begin{equation}
\dot{A}^{k}(x) =  \omega^{-1}(x)\left ( {\dot {\cal A}}^{k}(x) + \nabla^{k}({\cal A})X_{0} \right ) \omega(x)
\label{eq:dotA}
\end{equation}
It is then simple to verify that
\begin{equation}
\nabla_{k}(A) \dot {A}^{k}(x) = \omega^{-1}(x)\left (\nabla_{k}({\cal A}){\dot {\cal A}}^{k}
+ \nabla_{k}({\cal A})\nabla^{k}({\cal A})X_{0} \right ) \omega(x)
\label{eq:cond}
\end{equation}
Imposing the condition (\ref{eq:genCoul}) on the potential $A_{k}(x)$ is thus, according to Eq.     
(\ref{eq:cond}), equivalent to the following requirement,
\begin{equation}
\nabla_{k}({\cal A})\nabla^{k}({\cal A})X_{0} = - \nabla_{k}({\cal A}){\dot {\cal A}}^{k}  
\label{eq:ellipt}
\end{equation} 
But Eq. (\ref{eq:ellipt}) is a {\em linear elliptic partial differential equation} for the
Lie algebra valued (matrix valued) quantity $X_{0}$ when the general unconstrained potential
${\cal{A}}_{k}$ is considered given. Thus the existence and uniqueness of the gauge condition
(\ref {eq:genCoul}) is essentially equivalent to the question of existence and uniqueness of 
solutions $X_{0}$ to the linear elliptic partial differential equation (\ref{eq:ellipt}). 

I have so far been a little cavalier with respect to the question of the domain of the independent
variables ${\bf x} \in {\bf R}^{3}$ in Eq. (\ref{eq:ellipt}), where the time variable $x^{0}$ is a
parameter.

It is appropriate to consider the differential equation (\ref{eq:ellipt}) for any fixed $x^{0}$
in a finite simply connected domain $V$ with a smooth boundary $\partial V$, such as for instance 
a ball $B_{R}$ of radius $R$, centered at the origin of space coordinates, i.e.
\begin{equation}
B_{R}  = \left \{ {\bf x} | \mid {\bf x} \mid < R \right \}
\label{eq:domain}
\end{equation}
with a view of letting $V$ grow indefinitely (e.g. $R \rightarrow \infty$) later at some appropriate
stage in the developmemnt of the formalism.

In such a situation the {\em existence} of a solution $X_{0}$ to Eq. (\ref{eq:ellipt}) can not really
be cast in doubt, provided only the unconstrained gauge potential $\cal {A}$ satisfies some mild 
regularity conditions in $V$. The {\em uniqueness} of the solution $X_{0}$ is again related to the 
boundary conditions one wishes to impose on the quantity $X_{0}$ at the boundary $\partial V$.
 
We defer the discussion of the important question of possible boundary conditions till later,
and consider in stead how a solution $X_{0}$ of Eq. (\ref{eq:ellipt}) determines the gauge 
transformation $\omega(x)$, which is the object of prime interest in the present discussion. 
From the relation (\ref{eq:capX}) follows that
\begin{equation}
\dot{\omega}(x^{0}, {\bf x}) = igX_{0}(x^{0}, {\bf x})\omega (x^{0}, {\bf x})
\label{eq:solomega}
\end{equation}
where for convenience we have split the argument $x$ into its time ($=x^{0}$) and space ($={\bf x}$)
components.

But Eq. (\ref{eq:solomega}) has the immediate solution
\begin{equation}
\omega (x^{0}, {\bf x}) = \left [ T \exp (ig \int_{z^{0}}^{x^{0}} d\tau X_{0}(\tau, {\bf x})) \right ] \omega (z^{0}, {\bf x})
\label{eq:finomega}
\end{equation}
where $T$ stands for time ordering in the exponential integral, and $\omega (z^{0}, {\bf x})$
is an initial value, which contains an arbitrary dependence on the space variables ${\bf x}$. Thus,
apart from an inessential initial value, depending on the space coordinates ${\bf x}$ only, the
solution $X_{0}$ of the partial differential equation (\ref {eq:ellipt})  defines the gauge transform
$\omega(x)$, by means of which the gauge condition (\ref{eq:genCoul}) is attained, uniquely. The 
freedom of choosing any suitable initial value $\omega (z^{0}, {\bf x})$ in the relation 
(\ref{eq:finomega}) is a somewhat trivial ambiguity, and is not what is meant by a Gribov ambiguity. 
One can easily get rid of this type of residual gauge freedom by supplementing  the condition 
(\ref{eq:genCoul}) by a suitable additional gauge condition at the fixed initial time $z^{0}$. 
A possible additional condition of this kind is for instance the following,
\begin{equation}
x^{k}A_{k}(z^{0}, {\bf x}) = 0
\label{eq:radigauge}
\end{equation}
which then fixes the gauge completely, apart from constant gauge transformations, naturally.

This concludes the demonstration that (\ref{eq:genCoul}) {\em is} a {\em bona fide} gauge 
condition, which is attainable by means of a gauge transform $\omega(x)$, which is determined 
(uniquely modulo initial and boundary conditions) by the linear  elliptic partial differential 
equation  (\ref{eq:ellipt}) for the Lie algebra valued quantity $X_{0}$.

\section{The Gauss law constraints}

In this section I consider the field equations for the Yang-Mills potentials, and show that the 
equation of constraint among these equations, namely the non-Abelian Gauss law, gets expurgated 
of any time derivatives if one uses the generalised Coulomb gauge (\ref{eq:genCoul}), and in fact
reduces to a form which resembles the corresponding equation in electrodynamics  as closely as 
possible \cite{Weinberg}. 

The Yang-Mills action $S$, with a general potential ${\cal A}$, which is not 
supposed to satisfy any particular gauge condition, is as follows,
\begin{equation}
S = - \; \frac{1}{4}\; \int d^{4}x (G_{\mu\nu}({\cal A}), G^{\mu\nu}({\cal A}))
\label{eq:action}
\end{equation}
As is well known, requiring the action (\ref{eq:action}) to be stationary with respect to 
variations of {\em all} the potential components ${\cal A}_{\mu}$, considered as independent
quantities, yields the follwing field equations,
\begin{equation}
\nabla_{\nu}({\cal A}) G^{\mu\nu}({\cal A}) = 0
\label{eq:fieldeq}
\end{equation}
A perhaps more interesting system of equations would be obtained by coupling the gauge field to 
appropriate matter fields. Then the right hand side of Eq. (\ref{eq:fieldeq}) would be 
replaced by a covariantly conserved matter current. Such an addition is not absolutely essential 
for the questions pursued in this paper, and will therefore not be contemplated further here.
The non-Abelian Gauss law is obtained from the equations (\ref {eq:fieldeq}) for $\mu = 0$. 
Expressed in terms of the potential ${\cal A}$ the non-Abelian Gauss law reads as follows, 
\begin{equation}
\nabla_{k}({\cal A})\nabla^{k}({\cal A}) {\cal A}^{0} -  \nabla_{k}({\cal A}){\dot {\cal A}}^{k} = 0
\label{eq:nAGauss}
\end{equation}
A possible way of obtaining a canonical formalism for the Yang-Mills system, incorporating 
the constraint Eq. (\ref{eq:nAGauss}), is to use the gauge freedom to set 
${\cal A}^{0}$ equal to zero.  In this so-called Weyl gauge \cite{Weyl} ($A^{0} = 0$), the 
non-Abelian Gauss law (\ref {eq:nAGauss}) is the following,
\begin{equation}
\nabla_{k}(A){\dot A}^{k} = 0
\label{eq:WeylnAG}
\end {equation}
whereas the remaining equations of motion obtained from Eq. (\ref {eq:fieldeq}) are as follows,
\begin{equation}
{\dot A}^{k} = - G^{0k}, \; {\dot G}^{0k} = \nabla_{\ell}(A) G^{k\ell}(A)
\label{eq:Weqmo}
\end{equation}
The by now traditional way \cite{Jackiw} of understanding the equations (\ref {eq:WeylnAG})
and (\ref {eq:Weqmo}) as a canonical Hamiltonian system, is to first disregard the constraint
(\ref {eq:WeylnAG}), and to concentrate on finding a canonical Hamiltonian system of equations 
equivalent to the equations of motion (\ref {eq:Weqmo}) alone. This is essentially trivial, and
after quantisation one then incorporates the omitted constraint (\ref {eq:WeylnAG}) as a 
condition on the set of states in the quantum version of the theory. While the procedure outlined 
briefly above is in many ways very interesting, it has  not, to the best of my knowledge, yielded
deep insight into the confinement question in pure Yang-Mills theory, i.e. the question of the
glueball spectrum. The gauge condition (\ref {eq:genCoul}) proposed in this paper enables one to 
elucidate the canonical structure of pure Yang-Mills theory in a way which differs from the 
traditional Weyl gauge canonical formalism outlined above. 

Imposing the generalised Coulomb gauge condition (\ref {eq:genCoul}) one observes that the 
non-Abelian Gauss law (\ref {eq:nAGauss}) reduces to the following,
\begin{equation}
\nabla_{k}(A)\nabla^{k}(A) A^{0}(x^{0}, {\bf x}) = 0
\label{eq:genCnAG}
\end{equation}
The equation (\ref {eq:genCnAG}) resembles the corresponding equation in electrodynamics as 
closely as possible, and defines the quantity $ A^{0}$ as a functional of the space
components $A_{k}, k = 1,2,3$ only, unless a dependence of the time derivatives of the potential
components is introduced via the boundary conditions for the equation in question.

It should be noted that  equation (\ref {eq:genCnAG}) in the general case has nontrivial solutions, 
despite being a homogeneous equation. Solutions which are regular for finite $\mid {\bf x} \mid$, 
will in general have nontrivial asymptotic behaviour for large $\mid {\bf x} \mid$.
 
Actually, the more interesting situation obtains if the gauge field gets coupled to the appropriate 
matter fields, in which case the right hand side of Eq. (\ref {eq:genCnAG}) gets replaced by a
non-vanishing density. This generalisation may affect the possible asymptic behaviour of the
solution $A^{0}$ of the corresponding inhomogeneous version of Eq. (\ref {eq:genCnAG}).

Implementing the particular version of the non-Abelian Gauss law (\ref {eq:genCnAG}) straightforwardly,
as an equation determining the zero component $A^{0}$ of the gauge potential, when the space components
$A^{k}, k = 1,2,3$ are supposed to be given, enables one to obtain a canonical formulation of Yang-Mills 
theory in the gauge (\ref {eq:genCoul}) which is substantially different from the Weyl gauge treatment
briefly presented above. Details of this construction will be given in a forthcoming paper 
\cite{Chris 1}.

\section{Summary and Conclusions}   

In this paper I have shown that there is a gauge condition, Eq. (\ref {eq:genCoul}), in Yang-Mills
theory, which is a straightforward generalisation of the ordinary Coulomb gauge condition in
electrodynamics. This new gauge condition is neither more nor less than what is needed in order to 
reduce the non-Abelian Gauss law, in its general form given by Eq. (\ref {eq:nAGauss}),  to an 
equation not involving any time derivatives of combinations of potential components, namely Eq.
(\ref {eq:genCnAG}), in perfect analogy with the corresponding equation in the Coulomb gauge 
formulation of electrodynamics. The particular version of the non-Abelian Gauss law, Eq. 
(\ref {eq:genCnAG}), which is valid in the gauge (\ref{eq:genCoul}), is a (homogemeous) linear 
elliptic partial differential equation in the space variables ${\bf x}$, with time $x^{0}$ acting as 
a parameter. This partial differential equation determines the $A^{0}$-component of the gauge 
potential for given space components $A^{k}, k = 1,2,3$. The boundary conditions on the quantity 
$A^{0}$, i.e. the asymptotic behaviour of $A^{0}(x^{0}, {\bf x})$ for large $\mid {\bf x} \mid$, 
play an important role in the analysis of the solutions to the equation in question. 

The proof that the proposed gauge condition (\ref {eq:genCoul}) 
actually {\em is} a gauge condition,  relies on the existence of solutions of a linear 
elliptic partial differential equation, namely Eq. (\ref{eq:ellipt}), which is non-homogeneous, but 
otherwise of the very same form as the equation (\ref {eq:genCnAG}). There is a certain unification
operating here; both the new gauge condition (\ref {eq:genCoul}), and the corresponding particular
version of the non-Abelian Gauss law (\ref {eq:genCnAG}), require the elucidation of the conditions 
under which there exist solutions to a certain linear elliptic partial differential equation. This 
equation is in any case one of the basic ingredients in the Yang-Mills equations,  and requires such 
an analysis regardless of whether one is interested in the generalised Coulomb gauge condition 
(\ref{eq:genCoul}) or not.
 
To the extent that appropriate boundary conditions select an essentially unique solution to the 
linear elliptic partial differential equation in question, Eq. (\ref {eq:ellipt}), there are no 
Gribov ambiguities associated with the new generalised Coulomb gauge condition (\ref {eq:genCoul}) 
proposed here. The freedom of making gauge transformations depending on the space coordinates 
${\bf x}$ only, which is permitted by the new gauge condition (\ref{eq:genCoul}), can trivially be 
reduced to the freedom of making global, i.e. constant  gauge transformations only, by using a 
supplementary gauge condition such as Eq. (\ref {eq:radigauge}) above.  

Finally, it is possible to use the formulation of Yang-Mills theory in the new gauge 
(\ref {eq:genCoul}) proposed here, as a starting point in developing a canonical \cite{Chris 1}
formalism for Yang-Mills theory, which differs in substance from the traditional Weyl gauge 
treatment \cite{Jackiw}.

\end{document}